
\magnification=1200
\baselineskip=13pt
\tolerance=100000
\overfullrule=0pt

\rightline{UR-1344\ \ \ \ \ \ \ }
\rightline{ER-40685-793}

\bigskip

\baselineskip=20pt

\centerline{\bf TESTS OF INTEGRABILITY OF THE SUPERSYMMETRIC}
\centerline{\bf NONLINEAR SCHR\"ODINGER EQUATION}
\vskip .5in

\centerline{J. C. Brunelli}
\centerline{and}
\centerline{Ashok Das}
\centerline{Department of Physics and Astronomy}
\centerline{University of Rochester}
\centerline{Rochester, NY 14627}

\vskip .5in

\centerline{\bf \underbar{Abstract}}

\medskip

We apply various conventional tests of integrability to the supersymmetric
nonlinear Schr\"odinger equation.  We find that a matrix Lax pair exists
and that the system has the Painlev\'e property only for a particular
choice of the free parameters of the theory.  We also show that the second
Hamiltonian structure generalizes to superspace only for these values of
the parameters.  We are unable to construct a zero curvature formulation of
the equations based on OSp(2$|$1).  However, this attempt yields a
nonsupersymmetric fermionic generalization of the nonlinear Schr\"odinger
equation which appears to possess the Painlev\'e property.

\vfill\eject

\noindent {\bf I. Introduction:}

\medskip

The nonlinear Schr\"odinger equation is one of the most widely studied
1 + 1 dimensional integrable systems [1-3].
  In contrast, the supersymmetric
nonlinear Schr\"odinger (SNLS)
 equation has not drawn as much attention.  In fact,
it was only very recently that the supersymmetrization of the nonlinear
Schr\"odinger equation was proposed.  It was shown in ref. 4   through a
study of the prolongation structure [5]
 that there are two distinct supersymmetrizations -- one with a free
parameter -- which are integrable.  The structure of a supersymmetric
theory is, in general, more restrictive than its bosonic
counterpart [6,7].  We
were, therefore, quite puzzled by the appearance of a free parameter in the
supersymmetric nonlinear Schr\"odinger equation and decided to study this
system more systematically.  We find that various conventional tests of
integrability select out only a supersymmetric system without any free
parameter.

In sec. II, we briefly review the
 supersymmetric nonlinear Schr\"odinger equation
 both in components and in superspace.  This
also establishes our notations and conventions relative to
ref. 4.  In sec. III, we construct the matrix Lax pair for the
supersymmetric equation in superspace and show that this is possible only
if the supersymmetric equations have no free parameter.  In sec. IV, we
carry out the Painlev\'e analysis [8,9]
 for the supersymmetric equation and we can
construct a consistent set of solutions only for the case without any free
parameters as in sec. III.  In sec. V, we try to study the
 zero curvature formulation [10,11] of the supersymmetric system.  We
show that it is not possible to derive these equations from a zero
curvature condition based on OSp(2$|$1) [12,13].
  However, we can obtain a fermionic
generalization very close in form to the supersymmetric equation without
any free parameter.  This fermionic generalization, even though not
supersymmetric, seems to pass the Painlev\'e test and, therefore, we believe
that this represents a new integrable system.  In sec. VI, we analyze the
Hamiltonian structures for the supersymmetric nonlinear Schr\"odinger
equation and
show that a second Hamiltonian structure exists only for the case without
any free parameter. This is very similar to what happens in the analysis of
 the supersymmetric KdV equation.
 Finally, we present our conclusions in sec.~VII.

\medskip

\noindent {\bf II. SNLS Equations:}

\medskip

The nonlinear Schr\"odinger equation is described in terms of a complex
bosonic variable $q(x,t)$ and has the form
$$\eqalign{iq_t &= -q_{xx} + 2k (q^*q)q\cr
-iq^*_t &= -q^*_{xx} + 2k (q^*q)q^*\cr}\eqno(2.1)$$
Here the subscripts $t$ and $x$ refer to derivatives with respect to these
variables.  The parameter $k$ is real, arbitrary and its value can be set
to unity by suitably rescaling the dynamical variables $q(x,t)$ and
$q^*(x,t)$.  A simple dimensional analysis shows that we can assign the
following canonical dimensions to the various variables.
$$\eqalign{[x]&=-1\cr
[q] &= [q^*] = 1\cr}
\qquad \eqalign{[t] &= -2\cr
[k] &= 0\cr}\eqno(2.2)$$

The supersymmetric generalization of the nonlinear Schr\"odinger equation
was proposed in ref. 4.  In addition to the complex, bosonic variables
$q(x,t)$, it also involves complex, fermionic variables $\phi (x,t)$ and
has the form
$$\eqalign{iq_t &= -q_{xx} + 2k (q^*q)q + 2k(\alpha + \gamma)
q^* \phi_x \phi - 2k \alpha q \phi \phi^*_x\cr
&\qquad\qquad + 2k (2-2 \alpha - \gamma)q \phi_x \phi^* + 2k \gamma \phi
\phi^* q_x\cr
-iq_t^* &= -q_{xx}^* + 2k (q^*q)q^* + 2k(\alpha + \gamma)
q \phi_x^* \phi^* - 2k \alpha q^* \phi^* \phi_x\cr
&\qquad\qquad + 2k (2-2 \alpha - \gamma)q^* \phi_x^* \phi + 2k \gamma
\phi^* \phi q_x^*\cr
i\phi_t &= -\phi_{xx} + 2k \alpha q^* q  \phi
+ 2k (1- \alpha) q^2 \phi^* + 2k \gamma
\phi \phi^* \phi_x \cr
-i \phi^*_t &= -\phi^*_{xx}  + 2k \alpha q^* q \phi^*
+ 2k (1-\alpha )q^{*2} \phi + 2k \gamma \phi^* \phi \phi^*_x\cr}
\eqno(2.3)$$
Comparing with the corresponding equations in ref. 4
 it is easy to see that they are
related by $q \rightarrow \sqrt{2}\ q,\ q^* \rightarrow \sqrt{2} \ q^*$
with the identifications
$$\eqalign{c_1 &= 2k (\alpha + \gamma -1)\cr
c_2 &= 2k (\alpha -1)\cr}\eqno(2.4)$$
The set of equations (2.3) can be easily checked to be invariant under the
supersymmetry transformations
$$\eqalign{\delta q &= \epsilon \phi_x\cr
\delta \phi &= \epsilon q\cr}
\qquad \eqalign{\delta q^* &= \epsilon \phi^*_x\cr
\delta \phi^* &= \epsilon q^* \cr}\eqno(2.5)$$
where $\epsilon$ is a real, constant Grassmann parameter.

A little bit of analysis shows that we can assign the following canonical
dimensions to the various variables in Eq. (2.3).
$$\eqalign{[x] &= - 1 \qquad\qquad [t] = -2\cr
[q] &= [q^*] = 1\cr
[\phi ] &= [\phi^*] = {1 \over 2}\cr
[k] &= [\alpha] = [\gamma] = 0\cr}\eqno(2.6)$$
The equations (2.3) can be put in a manifestly supersymmetric form by
introducing a complex, fermionic superfield
$$\Phi (x,t, \theta) = \phi (x,t) + \theta q (x,t) \eqno(2.7)$$
Here $\theta$ is the real, Grassmann coordinate of the superspace and by
construction
$$[\theta ] = - {1 \over 2}\qquad \qquad [\Phi ] = {1 \over 2} \eqno(2.8)$$
Introducing the covariant derivative in superspace to be
$$D = {\partial \over \partial \theta} +
\theta \ {\partial \over \partial x} \eqno(2.9)$$
the equations in (2.3) can be written also as
$$\eqalign{i \Phi_t &= -D^4 \Phi + 2k \alpha (D \Phi^*) (D \Phi) \Phi
- 2k \gamma \Phi^* \Phi (D^2 \Phi ) + 2k (1 - \alpha) \Phi^*
(D \Phi)^2\cr
-i \Phi_t^* &= -D^4 \Phi^* + 2k \alpha (D \Phi) (D \Phi^*) \Phi^*
- 2k \gamma \Phi \Phi^* (D^2 \Phi^* ) + 2k (1 - \alpha) \Phi
(D \Phi^*)^2\cr}\eqno(2.10)$$
It is easy to see that this is the most general equation
 (with the appropriate phase invariance) in superspace
(note $[D] = {1 \over 2}$) which will reduce to the nonlinear Schr\"odinger
equation in the bosonic limit.  Even though the parameters $\alpha$,
$\gamma$ (and, therefore, $c_1$ and $c_2$) are arbitrary, it was shown in
ref. 4 that the system of equations (2.3) are integrable only if
$$c_1 = - 4 k \qquad \qquad c_2 = 0 \eqno(2.11)$$
or
$$c_1 = c = \ {\rm arbitrary} \qquad \qquad c_2 = 4k \eqno(2.12)$$
To further analyze integrability, we will construct the matrix Lax pair for
the system of equations (2.3) in the next section.

\medskip

\noindent {\bf III. Matrix Lax Pair:}

\medskip

The nonlinear Schr\"odinger equation, Eq. (2.1), can be written in terms of
a Lax pair where the Lax operators are 2 $\times$ 2 matrices.
Conventionally [14], it is well known that the Lax pair
$$L = \pmatrix{i(1+ \beta) {\partial \over \partial x} &q^*\cr
\noalign{\vskip 5pt}%
q & i (1 - \beta ){\partial \over \partial x}\cr}$$
\rightline{(3.1)}
$$B = \pmatrix{-i \beta {\partial^2 \over \partial x^2} + {i
q^* q \over 1 + \beta} &-q^*_x\cr
\noalign{\vskip 5pt}%
q_x & -ik {\partial^2 \over \partial x^2} - {i q^*q \over
1 - \beta}\cr}$$
where $k = {2 \over 1 - \beta^2}$, would give Eq. (2.1) from the Lax
equation
$$L_t = [L,B] \eqno(3.2)$$
It is quite straightforward to check that the Lax pair can be generalized
to include a dimensionful parameter, $\zeta$.  In fact, the pair
$$L = \pmatrix{(1+ \lambda) {\partial \over \partial x} + i \zeta &(1 -
\lambda^2)^{1/2} k^{1/2} q^*\cr
\noalign{\vskip 10pt}%
(1-\lambda^2)^{1/2} k^{1/2} q &(1-\lambda) {\partial \over
\partial x} + i \zeta\cr}$$
\rightline{(3.3)}
$$B = \pmatrix{-i \lambda {\partial^2 \over \partial x^2} + 2i \zeta^2
-ik (1- \lambda)q^*q & -i(1- \lambda^2)^{1/2} k^{1/2} q^*_x\cr
\noalign{\vskip 10pt}%
i(1- \lambda^2)^{1/2} k^{1/2}q_x &-i \lambda {\partial^2 \over
\partial x^2} + 2i \zeta^2 + ik (1+ \lambda)
q^*q\cr}$$
where $\lambda\  (\zeta)$ is an arbitrary dimensionless (dimensionful)
parameter also give Eq. (2.1) from the Lax Eq. (3.2).
Thus, there appears to be a certain degree of freedom in the choice of the
matrix Lax pair for the nonlinear Schr\"odinger equation.  However, we find
that neither of the forms of the Lax pair in Eqs. (3.1) and (3.3)
generalize to the supersymmetric case.

The most logical choice for the Lax pair, in the case of the supersymmetric
theory, would be in terms of the superfields.  To construct the Lax pair,
we note that
$$\eqalign{[L] &= 1 \qquad \qquad [B] = 2\cr
[\Phi ] &= [\Phi^*] = {1 \over 2} \cr}\eqno(3.4)$$
We write the most general ansatz, consistent with the canonical dimensions
in Eq. (3.4), for $L$ and $B$ in terms of the superfields $\Phi,$
 $\Phi^*$ and the covariant derivative $D$.  Requiring that the Lax
equation
$$L_t = [L,B]$$
gives the manifestly supersymmetric equations in Eq. (2.10), we find that
this is possible only if $\alpha = 1= -\gamma$.  In this case, the Lax pair
has the form

\medskip

$$L = \pmatrix{\lambda D^2 + i \zeta + \lambda k \Phi^* \Phi
&\lambda k^{1/2} (D \Phi^*)\cr
\noalign{\vskip 10pt}%
\lambda k^{1/2} (D \Phi) &\lambda D^2 + i \zeta -
\lambda k \Phi^* \Phi\cr}$$
\rightline{(3.5)}
$$B = \pmatrix{(i+\beta)D^4 + \beta k (D\Phi^*) (D\Phi) + 2 i \zeta^2
&\beta k^{1/2} (D^3 \Phi^*) + 2(i + \beta)k^{1/2} (D \Phi^*)D^2\cr
\qquad -ik ((D^2 \Phi^*)\Phi - \Phi^* (D^2 \Phi)) &\cr
\qquad + (i+ \beta) k\{(D^2 (\Phi^* \Phi)) + 2
\Phi^* \Phi D^2\} &\cr
\noalign{\vskip 10pt}%
(2i+\beta)k^{1/2} (D^3 \Phi) + 2(i+ \beta)k^{1/2} (D\Phi)
&(i+ \beta) D^4 + (2i+\beta)k (D \Phi^*) (D \Phi)\cr
&\qquad + 2i\zeta^2 +ik ((D^2 \Phi^*)\Phi - \Phi^* (D^2 \Phi)) \cr
&\qquad - (i+ \beta) k\{(D^2 (\Phi^* \Phi)) + 2
\Phi^* \Phi D^2\} &\cr}$$
where $\lambda$ and $\beta$ are arbitrary, dimensionless parameters and
$\zeta$ is a dimensionful arbitrary parameter.

It is believed that an integrable system can be written as a Lax equation
in terms of a pair of operators $L$ and $B$.  The very fact that a Lax form
of the equation exists only when $\alpha = 1= -\gamma$ would then seem to
indicate that the supersymmetric system is integrable only when there is no
free parameter.  (Incidentally, this choice of the parameters can be
 seen from Eq. (2.4) to imply that the system is integrable only for
 $c_1 = -2k$,
$c_2 = 0$.  This is very similar to Eq. (2.11) and we are hoping that there
is a misprint in ref. 4).  We also wish to point out that when $\alpha
= 1= - \gamma$, the supersymmetric equations have the form (omitting the
complex conjugate equations)
$$\eqalign{iq_t &= -q_{xx} + 2k (q^* q)q - 2k q \phi \phi^*_x
+ 2k q \phi_x \phi^* - 2k \phi \phi^* q_x\cr
i\phi_t &= -\phi_{xx} + 2k (q^*q)\phi - 2k \phi \phi^* \phi_x\cr}\eqno(3.6)
$$
While the set of equations in Eq. (2.3) are invariant under the global
phase transformations
$$\eqalign{q &\rightarrow e^{-i \epsilon}q \cr
\phi &\rightarrow e^{-i \epsilon} \phi\cr}
\qquad \eqalign{q^* &\rightarrow e^{i \epsilon} q^* \cr
\phi^* &\rightarrow e^{i \epsilon} \phi^*\cr} \eqno(3.7)$$
the set of equations in Eq. (3.6) have an invariance under a larger global
transformation
$$\eqalign{q &\rightarrow e^{-i \epsilon} q \cr
\phi &\rightarrow e^{-i \sigma} \phi\cr}
\qquad \eqalign{q^* &\rightarrow e^{i \epsilon} q^*\cr
\phi^* &\rightarrow e^{i \sigma} \phi^*\cr}\eqno(3.8)$$
To understand integrability further, we turn to a Painlev\'e analysis of the
system of equations in (2.3) in the next section.

\medskip

\noindent {\bf IV. Painlev\'e Analysis:}

\medskip

To carry out the Painlev\'e analysis [8,9] for the supersymmetric system in Eq.
(2.3) we treat $q$, $q^*=p$, $\phi$ and $\phi^* =\psi$
 as independent variables and
rewrite the equations as
$$\eqalign{iq_t &= -q_{xx} + 2k(pq)q + 2k (\alpha +\gamma)p \phi_x \phi
-2k \alpha q\phi \psi_x\cr
&\qquad\qquad + 2k (2-2 \alpha -\gamma)q \phi_x \psi +
2k \gamma \phi \psi q_x\cr
-ip_t &= -p_{xx} + 2k(pq)p + 2k (\alpha +\gamma)q \psi_x \psi
-2k \alpha p\psi \phi_x\cr
&\qquad\qquad + 2k (2-2 \alpha -\gamma)p \psi_x \phi +
2k \gamma \psi \phi p_x\cr
i\phi_t &= -\phi_{xx} + 2k \alpha (pq)\phi + 2k (1- \alpha)q^2 \psi
 + 2k \gamma  \phi \psi \phi_x\cr
-i \psi_t &= -\psi_{xx} + 2k\alpha (pq)\psi + 2k
 (1- \alpha)p^2 \phi
 + 2k \gamma  \psi \phi \psi_x\cr}\eqno(4.1)$$
Following the standard discussion [3], we choose a series expansion for the
independent variables of the form
$$\eqalign{q(x,t) &= \xi^{\beta_1} \sum^\infty_{j=0} q_j (t) \xi^j\cr
p(x,t) &= \xi^{\beta_2} \sum^\infty_{j=0} p_j (t) \xi^j\cr
\phi (x,t) &= \xi^{\delta_1} \sum^\infty_{j=0} \phi_j (t) \xi^j\cr
\psi (x,t) &= \xi^{\delta_2} \sum^\infty_{j=0} \psi_j (t) \xi^j\cr}
\eqno(4.2)$$
where we assume the singularity surface of the solutions to have the form
$$\xi (x,t) = x + f(t) = 0 \eqno(4.3)$$
We further assume that the zeroth order coefficients in the expansion in
Eq. (4.2) do not vanish, namely,
$$\eqalign{q_0 &\not= 0\cr
\phi_0 &\not= 0\cr}
\qquad
\eqalign{p_0 &\not= 0\cr
\psi_0 &\not= 0\cr}\eqno(4.4)$$
Note also that while the coefficients $q_j (t)$ and
$p_j (t)$ are bosonic, $\phi_j (t)$ and $\psi_j (t)$ are fermionic.

A leading order singularity analysis of Eq. (4.1) immediately gives
$$\beta_1 = \beta_2 = \delta_1 = \delta_2 = -1 \eqno(4.5)$$
Using these in Eq. (4.2) and substituting the series expansion into the
dynamical equations in (4.1) yields the recursion relations between the
coefficients.  These can be written compactly in the matrix notation as
$$M \chi = F \eqno(4.6)$$
where
$$\chi = \pmatrix{q_j(t)\cr
\noalign{\vskip 7pt}%
p_j (t)\cr
\noalign{\vskip 7pt}%
\phi_j (t)\cr
\noalign{\vskip 7pt}%
\psi_j (t)\cr} \eqno(4.7)$$
$$F = \pmatrix{F_j (t)\cr
\noalign{\vskip 7pt}%
\tilde F_j (t)\cr
\noalign{\vskip 7pt}%
G_j (t)\cr
\noalign{\vskip 7pt}%
\tilde G_j (t)\cr}\eqno(4.8)$$
and the supermatrix $M$ has the form
$$M = \pmatrix{A &D\cr
\noalign{\vskip 7pt}%
C &B\cr}\eqno(4.9)$$
and $A,\ B,\ C,\ D$ are 2 $\times$ 2 matrices.  The matrices $A$ and $B$ have
the form
$$A = \pmatrix{-(j-1)(j-2) + 4k q_0 p_0 &2k q^2_0 + 2k (\alpha
+ \gamma) \phi_1 \phi_0\cr
\quad + 2k (\alpha + (j-1) \gamma) \phi_1 \psi_0 &\cr
\quad + 2k (2 (\alpha -1) + j \gamma) \phi_0 \psi_1 &\cr
\noalign{\vskip 10pt}%
2k p^2_0 + 2k (\alpha + \gamma) \psi_1 \psi_0 & -(j-1)(j-2) +
4k q_0 p_0\cr
&\quad + 2k (\alpha + (j-1) \gamma)\psi_1 \phi_0\cr
&\quad +2k (2 (\alpha -1) +j \gamma)\psi_0 \phi_1\cr}\eqno(4.10)$$
and
$$B =\pmatrix{-(j-1)(j-2) + 2k \alpha q_0 p_0 & 2k (1-\alpha)q^2_0
 + 2k \gamma \phi_1 \phi_0\cr
\quad + 2k\gamma (1-j)\psi_0 \phi_1 - 2k \gamma j \psi_1 \phi_0 &\cr
\noalign{\vskip 10pt}%
2k (1-\alpha )p^2_0 + 2k \gamma \psi_1 \psi_0
 &-(j-1)(j-2) +2k \alpha q_0p_0\cr
&\quad + 2k\gamma (1-j)\phi_0 \psi_1 -
2k \gamma j \phi_1 \psi_0\cr}\eqno(4.11)$$

We note that Eq. (4.1) describes a set of four coupled second order
equations and, therefore, we expect eight arbitrary coefficients for a
unique solution of the system.  The arbitrariness will arise whenever the
matrix relations in Eq. (4.6) are not invertible.  For a supermatrix $M$ of
the form in Eq. (4.9), this happens when [15]
$$\det A = 0 \eqno(4.12)$$
and
$$\det B = 0 \eqno(4.13)$$
 \noindent From the forms
 of the matrices $A$ and $B$ in Eqs. (4.10) and (4.11) we
note that

\vfill\eject

$$\eqalignno{\det A &= \big( (j-1)(j-2) - 2k q_0 p_0 \big) \big( (j-1)(j-2) -
6 k q_0 p_0 \big)\cr
&\qquad + 2k (\alpha + \gamma -2)\big( (j-1)(j-2) - 4k q_0 p_0 \big)
(\psi_1 \phi_0 - \psi_0 \phi_1)\cr
&\qquad - 4k^2 (\alpha + \gamma)\big( q^2_0 \psi_1 \psi_0 + p^2_0 \phi_1 \phi_0
\big) &(4.14)\cr
\det B &= \big( (j-1)(j-2) +2k (1-2\alpha)q_0 p_0)
\big( (j-1)(j-2) -2k q_0 p_0 \big)\cr
&\quad + 2k\gamma \big((j-1)(j-2) - 2k \alpha q_0 p_0 \big)
\psi_1 \phi_0\cr
&\quad + 2k \gamma \big( (j-1)(j-2) -2k \alpha q_0 p_0 \big)
\phi_1 \phi_0 &(4.15)\cr
&\quad - 4k^2 \gamma (1 - \alpha) \big( q^2_0 \psi_1 \psi_0 +
p^2_0 \phi_1 \phi_0 \big) }$$
A low order analysis of the recursion relations gives
$$q_0 p_0 = {1 \over k} \eqno(4.16)$$
as in the bosonic nonlinear Schr\"odinger equation as well as
$$\psi_0 \propto \phi_0 \eqno(4.17)$$

Requiring $\det A$ and $\det B$ to vanish, we can determine the resonances.
 (These are the $j$-values at which arbitrary coefficients can arise.) From
the bosonic part of $\det A$ we obtain
$$j = -1, 0, 3, 4 \eqno(4.18)$$
while the vanishing of the bosonic part of $\det B$ gives
\itemitem{i)} $\alpha = {1 \over 2} \qquad j = 0, 1, 2, 3$ \hfill (4.19)

\itemitem{ii)} $\alpha = 1 \qquad j= 0,0,3,3$ \hfill (4.20)

\noindent The expressions in the determinants containing the fermionic
terms must also separately vanish and this can happen either through the
parameters $\alpha$ and $\gamma$ taking special values or the fermionic
coefficients having special relationships.  (Remember that Grassmann~var-

\vfill\eject

\noindent iables are nilpotent.)  However, the important conclusion of this
analysis is that the parameter $\alpha$ can only have two distinct values,
$\alpha = {1 \over 2}$ or 1.  We also note that the resonance values in Eq.
(4.18) correspond exactly to those of the nonlinear Schr\"odinger equation
(bosonic) and correspond to $j$-values where bosonic variables can become
arbitrary.  ($j=$$-1$ corresponds to the arbitrariness in the location of the
singularity surface.)  The other four resonances, (either Eq. (4.19) or
 (4.20)), therefore, would correspond to the $j$-~values where the fermionic
coefficients can become arbitrary.  A detailed analysis of the recursion
relations for $\alpha = {1 \over 2}$ yields inconsistencies.  We are,
therefore, left only with $\alpha = 1$.  Furthermore, we are able to obtain
consistent solutions in this case only for $\gamma = -1$.  As is clear from
Eqs. (4.18) and (4.20), there will be three arbitrary coefficients in this
case at $j=0$.  These correspond to
$$\phi_0 \quad , \quad \psi_0 (t) = \beta (t) \phi_0 \quad ,
\quad q_0 p_0 = {1 \over k} \eqno(4.21)$$
Here $\beta (t)$ is arbitrary.  Similarly, there are three arbitrary
coefficients at $j=3$ which can be identified with
$$\phi_3 (t) \quad , \quad \psi_3 (t) \quad {\rm and}\quad
(q_0 p_3 - q_3 p_0) \eqno(4.22)$$
The arbitrary bosonic coefficient at $j=4$ can be identified with
$$q_0 p_4 - q_4 p_0 \eqno(4.23)$$
To conclude this section, let us note that if we choose $\phi_3 \propto
\phi_0$ and $\psi_3 \propto \psi_0$, then some of the low order
coefficients are determined from this analysis to be
$$\eqalign{q_0 p_0 &= {1 \over k}\qquad \qquad \psi_0 = \beta (t) \phi_0\cr
\noalign{\vskip 4pt}%
\phi_1 &= - {i \over 2}\ \xi_t \phi_0 \qquad \qquad \psi_1
= {i \over 2}\ \xi_t \psi_0\cr
\noalign{\vskip 4pt}%
q_1 &= -{i \over 2}\ \xi_t q_0 - ik q_0 \phi_{0,t} \psi_0\cr
\noalign{\vskip 4pt}%
p_1 &= {i \over 2}\ \xi_t p_0 + ik p_0 \phi_{0,t} \psi_0\cr
\noalign{\vskip 4pt}%
\phi_2 &= {i \over 2}\ \phi_{0,t} - {1 \over 12}\ \xi^2_t \phi_0 -
{ik \over 3}\ q_{0,t}p_0 \phi_0\cr
\noalign{\vskip 4pt}%
\psi_2 &= -{i \over 2}\ \psi_{0,t} - {1 \over 12}\ \xi^2_t \psi_0 -
{ik \over 3}\ q_{0,t}p_0 \psi_0\cr
\noalign{\vskip 4pt}%
q_2 &= {i \over 6}\ q_{0,t} - {1 \over 12}\ \xi^2_t q_0 -
{k \over 6}\ \xi_t q_0 \phi_{0,t}\psi_0\cr
\noalign{\vskip 4pt}%
p_2 &= -{i \over 6}\ p_{0,t} - {1 \over 12}\ \xi^2_t p_0 -
{k \over 6}\ \xi_t p_0 \phi_{0,t} \psi_0\cr
\noalign{\vskip 4pt}%
q_3 p_0 + q_0 p_3 &= {1 \over 4k}\ \xi_{tt} - {1 \over 2}
\ \big( \phi_{0,tt} \psi_0 + \phi_{0,t} \psi_{0,t}\big)\cr}\eqno(4.24)$$
and so on.  We emphasize here that the Painlev\'e analysis seems to select
out $\alpha = 1 = - \gamma$ which is also the value we obtained in trying
to construct the matrix Lax pair in sec. III.

\medskip

\noindent {\bf V. Zero Curvature Formulation:}

\medskip

Much like the KdV equation, the nonlinear Schr\"odinger equation can also
be written as a zero curvature with potentials belonging to SL(2,{\bf R})
 (or SU(2)) [1,2].  In fact, the potentials
$$A_1 = \pmatrix{i \zeta & -ik^{1/2} q^*\cr
\noalign{\vskip 7pt}%
ik^{1/2} q &-i \zeta\cr}$$
\rightline{(5.1)}
$$A_0 = \pmatrix{-2i \zeta^2 - ik q^*q &-k^{1/2} q^*_x\cr
\noalign{\vskip 7pt}%
-k^{1/2}q_x &2i\zeta^2 + ik q^*q\cr}$$
give rise the field strength (curvature)
$$F_{01} =\partial_0 A_1 - \partial_1 A_0 + [A_0 , A_1] \eqno(5.2)$$
where $\partial_0 = {\partial \over \partial t}\ ,\ \partial_1 = {\partial
\over \partial x}$.  Requiring the field strength to vanish yields the
nonlinear Schr\"odinger equations of (2.1).  In trying to formulate the
supersymmetric nonlinear Schr\"odinger equation as a zero curvature, we
recall that the supersymmetric KdV equation can be formulated as a zero
curvature condition associated with the graded group OSp(2$|$1) [12,13].
OSp(2$|$1) has five generators and in components the zero curvature
condition has the form
$$\partial_0 A^I_1 - \partial_1 A^I_0 + f^{IJK} A^J_0 A^K_1 = 0
\eqno(5.3)$$
where $I,J,K = 1,2,3$ can be thought of as bosonic indices (the
corresponding generators satisfy commutation relations) while
$I,J,K = 4,5$ can be thought of as fermionic indices (the corresponding
generators satisfy anticommutation relations).  From Eq. (5.1), we see that
$$\eqalign{\big[A^1_1 \big] &= \big[ A^2_1 \big] = \big[ A^3_1 \big]
= 1 \cr
\big[ A^1_0\big] &= \big[ A^2_0 \big] = \big[ A^3_0 \big] = 2 \cr}
\eqno(5.4)$$
A simple dimensional analysis of Eq. (5.3) shows that
$$\eqalign{\big[ A^4_1 \big] &= \big[ A^5_1 \big] = 1\cr
\big[ A^4_0 \big] &= \big[ A^5_0 \big] = 2 \cr}\eqno(5.5)$$
Since the basic fermionic variables, $\phi$ and $\phi^*$, have canonical
dimension ${1 \over 2}$, the fermionic potentials must come multiplied by a
dimensional parameter of dimension ${1 \over 2}$.
 The supersymmetric equations, Eq. (2.3), on the
other hand, do not involve any dimensional parameter.  Therefore, the
equations, if they can be derived, must hold for any value of the
dimensional parameter.  However, we note that in the limit of the vanishing
dimensional parameter, the fermionic variables would drop out leading to an
inconsistency.  Thus, this simple argument shows that the supersymmetric
nonlinear Schr\"odinger equation cannot be formulated as a zero curvature
associated with Osp(2$|$1).  It is an open question as to whether it can be
expressed as a zero curvature associated with a different graded group.  If
it can be formulated as a zero curvature condition, through the standard
discussions, it can also be obtained from a self-duality condition [16].

Even though we have not succeeded in formulating the equations as a
zero curvature condition, we would like to point out the following.  Consider
the following potentials belonging to OSp(2$|$1).
$$A_1 = \pmatrix{k \phi^* \phi & -ik^{1/2} q^* & -i\lambda k^{1/2}
\phi^*\cr
\noalign{\vskip 7pt}%
ik^{1/2} q &-k \phi^* \phi &i \lambda k^{1/2} \phi\cr
\noalign{\vskip 7pt}%
i \lambda k^{1/2} \phi &i \lambda k^{1/2} \phi^* &0\cr}$$
\rightline{(5.6)}
$$A_0 = \pmatrix{-ik (q^*q + \phi_x \phi^* - \phi \phi^*_x )
&-k^{1/2}q^*_x &-\lambda k^{1/2} (\phi^*_x +
ik^{1/2} q^* \phi)\cr
\noalign{\vskip 7pt}%
-ik^{1/2}q_x &ik(q^*q + \phi_x \phi^* -\phi \phi^*_x )
&-\lambda k^{1/2} (\phi_x - ik^{1/2} \phi^* q)\cr
\noalign{\vskip 7pt}%
-\lambda k^{1/2} (\phi_x - ik^{1/2} \phi^* q) &\lambda k^{1/2} (\phi^*_x +
ik^{1/2} q^* \phi) &0\cr}$$
Here $[\lambda] = {1 \over 2}$.  The field strength
$$F_{01} = \partial_0 A_1 - \partial_1 A_0 + [A_0 , A_1 ]$$
can be constructed in a straightforward manner and requiring this to
vanish yields the equations
$$\eqalignno{iq_t &= - q_{xx} + 2k (q^*q)q - 2k q \phi \phi^*_x +
2k q \phi_x \phi^* - 2k q_x \phi \phi^*\cr
&\qquad + 2i \lambda k^{1/2} \phi_x \phi + 2 \lambda k q \phi^* \phi
 &(5.7)\cr
i \phi_t &= - \phi_{xx} + 2k (q^*q) \phi - 2 k
\phi \phi^* \phi_x &(5.8)\cr}$$
We recognize Eq. (5.8) as identical to the fermionic equation of the
supersymmetric equation for $\alpha = 1 = -\gamma$ given in Eq. (3.6).
Even the bosonic equation (5.7) is a generalization of the bosonic
equation in Eq. (3.6).  The new set of equations in Eqs. (5.7) and (5.8)
are no longer supersymmetric mainly because of the $\lambda$-dependent
terms.  However, one can think of them as a new fermionic extension of the
nonlinear Schr\"odinger equation similar to the earlier
construction [17,18].  A
preliminary analysis shows that these set of equations have the Painlev\'e
property and, therefore, are likely to be integrable.  It is tempting to
say that in the limit $\lambda \rightarrow 0$, the set of equations (5.7)
and (5.8) reduce to Eq. (3.6) and, therefore, we have a zero curvature
formulation for the special values $\alpha = 1 = - \gamma$.  However, as we
have argued when $\lambda \rightarrow 0$, fermionic variables drop out of
the potential and, therefore, in some sense it is an improper limit and the
zero curvature formulation of the supersymmetric nonlinear Schr\"odinger
equation remains an open problem.

\medskip

\noindent {\bf VI. The Hamiltonian structures:}

\medskip

Bosonic integrable systems such as the KdV system are known to have
biHamiltonian structures [19].  In fact, the biHamiltonian structures lead to
recursion operators which in turn lead to a nice geometric interpretation of
integrability in such systems [20,21].
  The nonlinear Schr\"odinger equation is
also known to have a biHamiltonian structure much like the KdV
system [1,19].  In
fact, if we define
$$H_1 = - \int dx \left( q^*_x q_x + k \left( q^*q \right)^2
\right) \eqno(6.1)$$
and
$$\left\{ q \left( x_1 , t \right) , q^* \left( x_2 , t \right)
\right\}_1 = i \delta \left( x_1 - x_2 \right) \eqno(6.2)$$
with all other Poisson brackets vanishing, then it is quite straightforward
to check that the nonlinear Schr\"odinger equation (Eq. (2.1)) can be
written as a Hamiltonian system,namely,
$$\eqalign{iq_t &= i \left\{ q , H_1 \right\}_1 \cr
-iq^*_t &= -i \left\{ q^* , H_1 \right\}_1 \cr}\eqno(6.3)$$
Conventionally, the Hamiltonian structure in Eq. (6.2) is known as the
first Hamiltonian structure.

Let us further note that we can also choose as a Hamiltonian for the system
$$H_2 = \int dx\ i \left( q^* q_x - q^*_x q \right) \eqno(6.4)$$
(Both $H_1$ and $H_2$ are conserved under the evolution of $q$ and
 $q^*$ given in Eq. (2.1).) If we now choose as the basic Poisson brackets
of the theory
$$\eqalign{\big\{ q \big( x_1 , t \big) , q\big( x_2 , t \big) \big\}_2
&= {k \over 2}\ q \big( x_1 ,t \big) q\big( x_2 ,t \big) \epsilon
\big( x_1 - x_2 \big)\cr
\noalign{\vskip 4pt}%
\big\{ q \big( x_1 , t \big) , q^* \big( x_2 , t \big) \big\}_2
&= {1 \over 2}\ \partial_{x_1} \delta \big( x_1 - x_2 \big) -
{k \over 2}\ q \big( x_1 ,t \big) q^* \big( x_2 ,t \big) \epsilon
\big( x_1 - x_2 \big)\cr
\noalign{\vskip 4pt}%
\big\{ q^* \big( x_1 , t \big) , q^* \big( x_2 , t \big) \big\}_2 &=
{k \over 2}\ q^* \big( x_1 ,t \big) q^* \big( x_2 ,t \big) \epsilon
\big( x_1 - x_2 \big)\cr}\eqno(6.5)$$
where $\epsilon (x_1 - x_2 )$ is the antisymmetric step function, then it
is straightforward to show that the nonlinear Schr\"odinger equation can be
written as a Hamiltonian system
$$\eqalign{iq_t &= i \big\{ q , H_2 \big\}_2 \cr
-iq^*_t &= - i \big\{ q^* , H_2 \big\}_2\cr}\eqno(6.6)$$
In other words, nonlinear Schr\"odinger equation is a biHamiltonian system
and Eq. (6.5) describes the second Hamiltonian structure of the theory.

It is known (mainly from the study of the supersymmetric KdV system) that
the two Hamiltonian structures of a bosonic integrable system do not
generalize to the supersymmetric
case [6,22,23].  In fact, for KdV, the first
Hamiltonian structure generalizes to superspace in the case of
 a supersymmetric theory that is not
integrable while it is the second Hamiltonian structure that generalizes to
the correct integrable supersymmetric theory [6].  While this is not a test of
integrability, it would be interesting to study the analogous situation for
the supersymmetric nonlinear Schr\"odinger equation.

Note that $H_1$ can be generalized readily to superspace as
$$\eqalign{H_1 = \int d \mu \bigg[ &{1 \over 2} \big( (D^3 \Phi^*)( D^2
\Phi) + (D^3 \Phi) (D^2 \Phi^*) \big) + {k \over 2}
((D \Phi^* )^2 \Phi (D \Phi)\cr
&+ (D \Phi)^2 \Phi^* (D \Phi^*)) + c\  \Phi^* \Phi ((D^2 \Phi)(D\Phi^*) -
(D^2 \Phi^*)(D \Phi))\bigg]\cr}\eqno(6.7)$$
Here \lq\lq $c$" is an arbitrary parameter and $d \mu = dx d\theta$,
 $\int d \theta = 0$ and $\int \theta d \theta = 1$.  (This
Hamiltonian reduces to Eq. (6.1) when the fermions are set equal to zero.)
Similarly, the first Hamiltonian structure in Eq. (6.2)  can be generalized
to superspace as
$$\left\{ \Phi \left( x_1 , \theta_1 , t \right), \Phi^* \left( x_2 ,
\theta_2 , t \right) \right\}_1 = -i D^{-1}_1 \Delta_{12} =
 -i D^{-1}_2 \Delta_{12} \eqno(6.8)$$
with all other Poisson brackets vanishing. Here
$$D_1 = {\partial \over \partial \theta_1} + \theta_1\ {\partial
\over \partial x_1} \eqno(6.9)$$
and
$$D^{-1}_1 = \partial^{-1}_{x_1} D_1 \eqno(6.10)$$
with
$$\Delta_{12} = \delta \left( x_1 - x_2 \right) \delta \left( \theta_1
- \theta_2 \right) = \delta \left( x_1 - x_2 \right)
\left( \theta_1 - \theta_2 \right) \eqno(6.11)$$
Requiring the superspace equations (see Eq. (2.10)) to be Hamiltonian of
the form
$$\eqalign{i \Phi_t &= i \big\{ \Phi , H_1 \big\}_1\cr
-i \Phi^*_t &= -i \big\{ \Phi^* , H_1 \big\}_1\cr}\eqno(6.12)$$
yields
$$c = -1 \qquad \gamma = 0 \qquad \alpha = 1 \eqno(6.13)$$
We see that the generalization of the first Hamiltonian structure selects
out a set of values
 of the parameters which does not satisfy the Painlev\'e property
much like the supersymmetric KdV case [6].

The generalization of the Hamiltonian in Eq. (6.4) to superspace is simpler
and has the form
$$H_2 = \int d\mu \ i \left( \left( D^3 \Phi^*
\right) \Phi - \left( D^3 \Phi \right) \Phi^* \right) \eqno(6.14)$$
The generalization of the second Hamiltonian structure in Eq. (6.5) needs a
little bit of work but can be determined to be
$$\eqalign{\big\{ \Phi \big( x_1 , \theta_1 , t \big) , \Phi \big( x_2
, \theta_2 , t \big) \big\}_2 &= k \Phi \big( x_1 , \theta_1 , t)
 \Phi \big( x_2 , \theta_2 , t\big) D^{-1}_1
\Delta_{12}\cr
\noalign{\vskip 4pt}%
\big\{ \Phi \big( x_1 , \theta_1 , t \big) , \Phi^* \big( x_2
, \theta_2 , t \big) \big\}_2 &= -{1 \over 2}\  D_1
\Delta_{12} - k \Phi \big( x_1 , \theta_1 ,
t \big)  \Phi^* \big( x_2 , \theta_2 , t\big) D^{-1}_1
\Delta_{12}\cr
\noalign{\vskip 4pt}%
\big\{ \Phi^* \big( x_1 , \theta_1 , t \big) , \Phi^* \big( x_2
, \theta_2 , t \big) \big\}_2 &= k \Phi^* \big( x_1 , \theta_1 ,
t \big) \Phi^* \big( x_2 , \theta_2 , t\big) D^{-1}_1
\Delta_{12}\cr}\eqno(6.15)$$
Requiring that the superspace equations be Hamiltonian with respect to Eqs.
(6.14) and (6.15), namely,
$$\eqalign{i \Phi_t &= i \big\{ \Phi , H_2 \big\}_2\cr
-i \Phi^*_t &= -i \big\{ \Phi^* , H_2 \big\}_2\cr}\eqno(6.16)$$
yields
$$\eqalign{i \Phi_t &= - D^4 \Phi + 2 k  (D \Phi^*)
(D \Phi) \Phi + 2k \Phi^* \Phi (D^2 \Phi)\cr
-i \Phi_t^* &= - D^4 \Phi^* + 2 k  (D \Phi)
(D \Phi^*) \Phi^* + 2k \Phi \Phi^* (D^2 \Phi^*)\cr}\eqno(6.17)$$
Comparing with Eq. (2.10), we note that this selects out the values of the
parameters to be
$$\alpha = 1 = - \gamma \eqno(6.18)$$
These are, of course, the values for which the Painlev\'e analysis goes
through.  Thus, once again, we see, as in the case of the supersymmetric KdV
equation, that the generalization of the second Hamiltonian structure
selects out the values of the parameters that are consistent with the
Painlev\'e analysis.

\medskip

\noindent {\bf VII. Conclusion:}

\medskip

We have studied the supersymmetric nonlinear Schr\"odinger equation
systematically.  We have  shown
 that various conventional tests of integrability
select out a theory where there is no arbitrariness in the parameters.
(This is consistent with one set of values Eq. (2.11) in ref. 4 assuming a
misprint.)  We have not succeeded in obtaining a zero curvature formulation
of this system based on the group OSp(2$|$1).  However, we obtain a
fermionic generalization of the theory which also appears to possess the
Painlev\'e property.  We have examined the generalization of the Hamiltonian
structures to superspace.  We find, much like the supersymmetric KdV
equation, that the generalization of the first Hamiltonian structure
selects out a set of values for the parameters which do not satisfy the
Painlev\'e property whereas the generalization of the second Hamiltonian
structure yields values of the parameter consistent with the Painlev\'e
analysis.  In light of our analysis, the second set of values of the
parameters containing a free parameter (see Eq. (2.12)) obtained in ref.
 4 from the study of the prolongation structure remains a puzzle.

This work was supported by U.S. Department of Energy Grant No.
DE-FG-02-91ER40685.  One of us (J.C.B.) would like to thank CNPq, Brazil
for financial support.

\vfill\eject

\noindent {\bf References:}

\medskip

\item{1.} L. D. Faddeev and L. A. Takhtajan, Hamiltonian methods in the
theory of solitons (Springer, Berlin, 1987).

\item{2.} A. Das, Integrable models (World Scientific, Singapore, 1989).

\item{3.} M. J. Ablowitz and P. A. Clarkson, Solitons, nonlinear evolution
equations and inverse scattering (Cambridge, New York, 1991).

\item{4.} G. H. M. Roelofs and P. H. M. Kersten, J. Math. Phys.
{\bf 33}, 2185 (1992).

\item{5.} F. B. Estabrook and W. D. Wahlquist, J. Math. Phys. {\bf 17},
1293 (1976).

\item{6.} P. Mathieu, J. Math. Phys. {\bf 29}, 2499 (1988).

\item{7.} A. Das, W-J. Huang and S. Roy, Phys. Lett. {\bf 157A}, 113
 (1991).

\item{8.} M. J. Ablowitz, A. Ramani and H. Segur, J. Math. Phys.
{\bf 21}, 715 (1980).

\item{9.} J. Weiss, M. Tabor and G. Carnevale, J. Math. Phys. {\bf 24}, 522
(1984).

\item{10.} S. S. Chern and C. K. Peng, Manuscr. Math. {\bf 28}, 207
(1979).

\item{11.} M. Crampin, F. A. Pirani and D. C. Robinson, Lett. Math. Phys.
{\bf 2}, 15 (1977).

\item{12.} A. Das and S. Roy, J. Math. Phys. {\bf 31}, 2145 (1990).

\item{13.} A. Das, W-J. Huang and S. Roy, Int. J. Mod. Phys. {\bf 15A},
3447 (1992).

\item{14.} V. E. Zakharov and A. B. Shabat, Sov. Phys. JETP {\bf 34}, 62
 (1972).

\item{15.} See for example, \lq\lq Group Theory in Physics", vol. 3 by J.
F. Cornwell, Academic Press (1989).

\item{16.} A. Das and C. A. P. Galv\~ao, Mod. Phys. Lett. {\bf A8}, 1399
(1993);
 J. C. Brunelli and A. Das, Davey-Stewartson equation from a zero
curvature and a self-duality condition, University of  Rochester preprint
UR-1332 (1993) (also hep-th/9312070) and references therein.

\item{17.} P. P. Kulish, ICTP, Trieste preprint IC/85/39, 1985.

\item{18.} R. A. Chowdhury and M. Naskar, J. Math. Phys. {\bf 28},
 1809 (1987).

\item{19.} F. Magri, J. Math. Phys. {\bf 19}, 1156 (1978).

\item{20.} S. Okubo and A. Das, Phys. Lett. {\bf 209B}, 311 (1988).

\item{21.} A. Das and S. Okubo, Ann. Phys. (N.Y.) {\bf 190}, 215 (1989).

\item{22.} W. Oevel and Z. Popowicz, Comm. Math. Phys. {\bf 139}, 441
(1991).

\item{23.} J. M. Figueroa-O'Farrill, J. Mas and E. Ramos, Leuven preprint
KUL-TF-91/19 (1991).

\end